\documentclass[preprintnumbers,amsmath,amssymb]{revtex4}
\usepackage{graphicx,color}
\begin{document}

\begin{center}
{\Large{\bf New interpretation of matter-antimatter asymmetry
based on branes and possible observational consequences}}
\end{center}

\vspace{0.5cm}

\centerline{Rong-Gen Cai$^1$, Tong Li$^2$, Xue-Qian Li$^2$ and Xun
Wang$^2$}

1. Institute of Theoretical physics, Chinese Academy of Sciences,
Beijing 100080

2. Department of Physics, Nankai University, Tianjin 300071

\vspace{1cm}

\begin{center}
\begin{minipage}{13cm}

\noindent Abstract:\\

Motivated by the AMS project, we assume that after the Big Bang or
inflation epoch, antimatter was repelled onto one brane which is
separated from our brane where all the observational matter
resides. It is suggested that CP may be spontaneously broken, the
two branes would correspond to ground states for matter and
antimatter respectively. Generally a complex scalar field which is
responsible for the spontaneous CP violation, exists in the space
between the branes and causes a repulsive force against the
gravitation. A possible potential barrier prevents the
mater(antimatter) particles to enter the space between two branes.
However, by the quantum tunnelling, a sizable anti-matter flux may
come to our brane. In this work by considering two possible
models, i.e. the naive flat space-time and Randall-Sundrum models
and using the observational data on the visible matter in our
universe as inputs, we derive the antimatter flux which would be
observed by the AMS detector.

\end{minipage}
\end{center}

\vspace{1cm}

\section{Introduction}

One of the major tasks of the modern particle-cosmology is to
explore a reasonable interpretation of the observed
matter-antimatter asymmetry in our universe\cite{antimatter}. The
widely acceptable picture is that there must exist three key
factors, namely, the CP violation\cite{CP1}, baryon number
non-invariance and existence of a stage out of equilibrium.
However, in the Standard Model (SM), the CP violation induced by
the non-zero CKM CP phase is not large enough to meet the
requirement\cite{CP2}. Therefore, either there is new physics
beyond SM which can cause larger CP violation\cite{CP3}, or exist
other mechanisms or picture which result in the observational
matter asymmetry.

One alternative interpretation was proposed that the antimatter
was repelled to other corners of the universe and in our part of
the universe, only matter resides. Thus definitely, the antimatter
would fly to our part and there is a possibility to be observed
before it annihilates with the regular matter particles. The
Alpha-Magnetic-Spectrometer project is set to observe the
anti-helium flux\cite{AMS1,AMS2,AMS}. On the theoretical aspect,
some authors have discussed existence of antimatter regions and
possible flux to our detector\cite{Khlopov}. They consider large
domains in the universe generated due to inflation, which then
convert into antimatter regions. The evolution of the antimatter
regions may result in an anti-star globular cluster. It is
interesting to note that the CP may be violated
spontaneously\cite{T.D.Lee,Weinberg}. In ref. \cite{Khlopov}, the
authors also suggest that separation of matter and antimatter is
caused by such spontaneous CP violation.

In this work, we propose another possible picture based on the
brane physics. Suppose that after the Big Bang or inflation epoch,
CP symmetry is spontaneously broken and  the two branes correspond
to different ground states of matter and antimatter respectively.
By the end of this phase transition, matter and antimatter begun
to reside on different branes. Following Goldberger and Wise
\cite{Wise}, we introduce a complex scalar fields which is
responsible for the spontaneous CP symmetry breaking. That is a
scalar field which only applies to the extra dimension and is
different from that in the standard model in our four-dimension
spacetime. The vacuum expectation values (VEV) may be CP-phase
dependent, so that the two branes correspond to different VEV's
and then accommodate matter and antimatter respectively.

According to the general theory of the brane physics, the gauge
fields are confined on each brane, but only the gravitational
field can extend to the extra dimension(s). The matter and
antimatter attract each other via gravitational force. To oppose
the gravitational attraction which may lead to a collision of the
two branes to destroy the universe, the scalar field can cause a
Casimir effect. The Casimir effect which can be calculated in the
qunatum field theory, results in a repulsive force against the
gravitational attraction. We find that the repulsive force caused
by the Casimir effect is much stronger than the gravitational
force as long as the separation of the two branes is small. Thus
it may cause a cosmological consequence, that at the early epoch
after the Big Bang, the two branes were closer, but they have been
repelled from each other and the trend would continue till some
day the two branes are sufficiently separated and then the
gravitational force observed in our four-dimensional spacetime
would obviously deviate from the Newton's universal gravitational
law. The picture seems to cause an unstable system. In the work
\cite{Wise}, the authors also suggested existence of a scalar
field which has different vacuum expectation values at two branes
(in their work, the other brane is empty) and a positive potential
which leads to a repulsive force between the two branes is
resulted in. The force can balance the gravitational force between
the branes as it is applied to our picture, but it depends on the
difference of the expectation values on the two branes. The
Casimir repulsive force is an alternative possibility.

Moreover, we suppose that there is a potential barrier at the
boundary of each brane which is similar to the surface tension for
a water membrane. The barrier prevents the matter or antimatter to
enter the space between the two branes and jump from one brane to
another. In this work, we describe the barrier by a delta
function, i.e.
$$V(\xi)=a\delta(\xi)+a\delta(\xi- \pi r_c),$$
where $a$ is a dimensionless parameter to be determined and $\pi
r_c$ is the separation between the two branes at present.

The antimatter may traverse across the gap between the two branes
via the quantum tunnelling. Thus an antimatter flux which has
already come in our universe, may freely propagate in our brane,
i.e. our matter world, until it annihilates with regular matter.
Since the matter density in our universe is dilute, as the first
order of approximation, we ignore its possible annihilation with
matter before the flux reaches our detector. Thus the AMS may
detect such antimatter flux and the measurement can provide us
some detailed information about the antimatter world. As
suggested, the AMS measures the flux of anti-helium from the
antimatter world. It is reasonable to suppose that the abundance
of anti-helium in the anti-matter world is the same as that of
helium in our matter-world, and then we can  estimate its flux.

We study the detection possibility in the naive flat-spacetime
model and the Randall-Sundrum model whose metric tensor is
suggested by the authors of \cite{RS}. In the Randall-Sundrum
model the "compactification radius" between two branes $r_c$ is
determined by solving the eletroweak hierarchy problem. Instead,
we let $r_c$ be a free parameters which must be much smaller than
1 mm for the observation of gravitational law and numerically
evaluate the flux of antimatter in our universe. However, we will
show that in the future universe the two branes will be repelled
away from each other by the Casimir force and finally the
gravitational balances the Casimir force and then $r_c$ would
reach its maximum and the observational Newton's law will be
obviously different from the present form. Indeed, we do not
expect to predict very accurate value for the antimatter flux, but
gain important information about such flux while the future AMS
experiment will help to fix concerned parameters.

After this introduction, we formulate the Casimir effect of the
scalar field and discuss its consequence. Then we derive the
Schr\"odinger equation for the fifth dimension in the
non-relativistic approximation and in the next section, we
evaluate the antimatter flux which penetrates the potential
barriers to reach our AMS detector. By the astronomical data we
roughly estimate the antimatter flux which can be captured by the
AMS detector. In the last section, we make more discussions and
draw our conclusion.

\section{Interaction between the two branes}

\subsection{The gravitational attraction between the two branes}

Different from the regular brane scenario where one brane is empty
and the normal matter resides on another, both branes are occupied
by massive particles and as the gravitational force line can cross
the fifth dimension, the two branes attract each other. Thus, let
us first estimate the gravitational attraction between the branes.

Considering a three-dimensional area $S$ on the brane where matter
uniformly distributes, the gravitational filed strength $E$ can be
derived in terms of the Gauss' law in four-dimension. The mass
density of the matter in our universe (anti-matter in the
anti-world) is $\rho$.

By the Gauss' law, we have
\begin{eqnarray}
{2\over G_5}ES&=&\rho S\nonumber\\
E&=&{G_5\rho\over 2},
\end{eqnarray}
where $G_5$ is the five-dimensional gravitational constant, whose
relation with four-dimensional gravitational constant $G_4$ is
basically $G_5=2r_cG_4$\cite{G45_1,G45_2}.

Thus the gravitational force density between the two branes reads
as
\begin{eqnarray}
f&=&E\rho=G_4r_c\rho^2.
\end{eqnarray}

\subsection{A possible Casimir force}

To balance the gravitational force between two branes, following
the literature, it is supposed that a scalar field exists between
the two branes and due to its existence there is a Casimir effect.
Under the periodic boundary condition, the Casimir energy density
induced by a massless scalar field is given
as\cite{Casimir1,Casimir2}
\begin{eqnarray}
V^P&=&{1\over 2}\sum^\infty_{n=-\infty}\int{d^4k\over
(2\pi)^4}\text{ln}(k^2+{n^2\over r_c^2})\nonumber \\
&=&-{1\over 2}{\partial \over \partial
s}|_{s=0}\left[\sum^\infty_{n=-\infty}\int{d^4k\over
(2\pi)^4}(k^2+{n^2\over r_c^2})^{-s}\right]\nonumber \\
&=&-{1\over 2}{\partial \over \partial
s}|_{s=0}\left[\sum^\infty_{n=-\infty}\int{d^4k\over
(2\pi)^4}{1\over \Gamma(s)}\int^\infty_0dte^{-(k^2+{n^2\over
r_c^2})t}t^{s-1}\right]\nonumber \\
&=&-{\partial \over \partial s}|_{s=0}\left[{\pi^2\over
(2\pi)^4r_c^4}{\Gamma(s-2)\over \Gamma(s)}(
r_c)^{2s}\zeta(2s-4)\right]\nonumber \\
&=&-{\pi^2\over (2\pi)^4r_c^4}\zeta'(-4)\nonumber \\
&=&-{\pi^2\over (2\pi)^4r_c^4}{3\over 4\pi^4}\zeta(5)\nonumber \\
&\simeq&-{3.102\over 64\pi^6r_c^4},
\end{eqnarray}
where $\zeta(5)\simeq1.034$.

The Casimir force density is
\begin{eqnarray}
F^P&=&-{\partial \over \partial r_c}V^P\nonumber \\
&=&-{12.408\over 64\pi^6r_c^5}.
\end{eqnarray}
That is an attractive force density and cannot play a role to
oppose the gravitational force. By contraries, if the boundary
condition is anti-periodic, one has the Casimir energy density as
\begin{eqnarray}
V^A&=&-{15\over 16}V^P\nonumber \\
&=&{3.102\over 64\pi^6r_c^4}{15\over 16},
\end{eqnarray}
and a repulsive Casimir force density
\begin{eqnarray}
F^A&=&{46.53\over 256\pi^6r_c^5}
\end{eqnarray}
is resulted in.

By the data, one can notice that for a small separation between
the two branes, i.e. the distance must be smaller than 1 mm
requested by the observation of gravitational law, the repulsive
Casimir force is larger than the attractive force between the two
branes. One can conjecture that at the early epoch of the
universe, the two branes were close to each other, and just due to
the repulsive force, the two branes gradually are repelled away
from each other and will continue to be separated further till one
typical distance which is about $10^5$ m, the Casimir force would
balance the gravitational force and then the observational
gravitational law  definitely deviates from the Newton's law, and
becomes\cite{G45_2}
\begin{equation}
V_4=-{G_4M\over r}(1+(n+1)e^{-\sqrt n r/R}), \label{newton}
\end{equation}
where $G_4$ is the four-dimension universal gravitational
constant, $n$ is the number of extra dimensions and $R$ is a
typical distance in the extra dimension.

The authors of Ref.~\cite{Wise} introduced an extra scalar field
and an interaction on the two branes in the Randall-Sundrum
scenario. The interaction of the scalar field between two branes
yields an effective four-dimensional potential for $r_c$. Then the
potential can help to stabilize $r_c$. The repulsive force caused
by the Casimir effects is another possibility.

\section{Transition rates of the anti-matter flux}

To obtain the transition rate of the anti-matter, one needs to
establish a Schr\"{o}dinger equation along the fifth dimension.
The form of the five-dimension Schr\"{o}dinger equation depends on
the metric for any concerned brane model. Below, we choose two
models, namely the naive flat space-time and
R-S(Randall-Sundrum)\cite{RS} metrics which are intensively
discussed in literature as examples to demonstrate how to evaluate
the anti-matter flux which would be observed by the AMS.

\subsection{Naive flat space-time}
We first discuss the simplest model, the naive flat space-time.
The metric for naive flat five-dimensional space-time is given as
\begin{equation}
ds^2 =\eta_{\mu\nu}dx^\mu dx^\nu+ d\xi^2,
\end{equation}
where $\eta_{\mu\nu}$ is the metric for a four-dimensional
Minkowski spacetime.  Substituting the metric into the
five-dimensional Klein-Gordon  equation, one has
\begin{eqnarray}
-({\partial\over
\partial t})^2\Psi+
({\partial\over
\partial \vec{x}})^2\Psi+({\partial\over
\partial \xi})^2\Psi-m^2\Psi&=&0.\label{formula2}
\end{eqnarray}
Decomposing the wavefunction $\psi$ into a product form
\begin{eqnarray}
\Psi&=&e^{i\vec{k}\cdot\vec{x}}\psi(\xi,t),
\end{eqnarray}
and substituting it into eq.(\ref{formula2}), one can eventually
obtain an equation which only contains differentiation of $\psi$
with respect to the fifth dimension variable $\xi$ and time $t$,
\begin{eqnarray}
\left[({\partial\over
\partial\xi})^2-({\partial\over \partial t})^2-(m^2+\vec{k}^2)\right]\psi(\xi,t)
&=&0\label{KG1}
\end{eqnarray}
It is noted that we ignore the regular interactions among the
particles on the branes. Then taking the non-relativistic
approximation,
\begin{eqnarray}
\psi(\xi,t)&=&\varphi(\xi,t)e^{-imt}\nonumber \\
-({\partial \over \partial t})^2\psi&\simeq&\left[2im{\partial
\varphi\over \partial t}+m^2\varphi\right]e^{-imt}
\end{eqnarray}
and substituting it into eq.(\ref{KG1}), we get
\begin{eqnarray}
2mi{\partial \varphi\over \partial t}+({\partial \over
\partial
\xi})^2\varphi+\left(-\vec{k}^2\right)\varphi&=&0,
\end{eqnarray}
namely, it is
\begin{eqnarray}
-{1\over 2m}({\partial \over \partial
\xi})^2\varphi+\left({\vec{k}^2\over
2m}\right)\varphi&=&E\varphi.\label{ADD}
\end{eqnarray}

After introducing two $\delta$ potentials at the surfaces of the
two branes and through a simple manipulation one has the
Schr\"odinger equation along the fifth dimension with an effective
potential and corresponding boundary conditions as
\begin{eqnarray}
-{1\over 2m}({\partial \over \partial
\xi})^2\varphi+\left({\vec{k}^2\over
2m}+a\delta(\xi)+a\delta(\xi-\pi
r_c)\right)\varphi&=&E\varphi.\label{ADD1}
\end{eqnarray}

At the anti-world brane, the solution of eq.(\ref{ADD1}) is:
\begin{eqnarray}
\varphi(\xi)&=&e^{i\alpha \xi}+R_1e^{-i\alpha
\xi}\;\;\;\;\;\;\;\;\;\;\xi<0,\nonumber \\
\varphi(\xi)&=&S_1e^{i\alpha \xi}+R_2e^{-i\alpha
\xi}\;\;\;\;\;\;\xi>0.
\end{eqnarray}
At our brane, the solution of eq.(\ref{ADD1}) is:
\begin{eqnarray}
\varphi(\xi)&=&S_1e^{i\alpha \xi}+R_2e^{-i\alpha
\xi}\;\;\;\;\;\;\xi<\pi r_c,\nonumber \\
\varphi(\xi)&=&S_2e^{i\alpha
\xi}\;\;\;\;\;\;\;\;\;\;\;\;\;\;\;\;\;\;\;\;\;\;\;\xi>\pi r_c.
\end{eqnarray}
where $\alpha$ is an eigenvalue for eq.(\ref{ADD}) as
$\alpha=\sqrt{2mE-\vec{k}^2}$.

The boundary conditions on the brane surfaces at $\xi=0$ and
$\xi=\pi r_c$ respectively demand
\begin{eqnarray}
\left(\varphi'(0^+)-\varphi'(0^-)\right)&=&2ma\varphi(0),\nonumber \\
\left(\varphi'((\pi r_c)^+)-\varphi'((\pi
r_c)^-)\right)&=&2ma\varphi(\pi r_c),
\end{eqnarray}
we can find the barrier penetration rate $T=|S_2|^2$ as
\begin{eqnarray}
T&=&|S_2|^2={(2\alpha)^2\over \left(4am+{2a^2m^2\sin(2\alpha\pi
r_c)\over \alpha}\right)^2+\left(2\alpha+{2a^2m^2(\cos(2\alpha\pi
r_c)-1)\over \alpha}\right)^2}.
\end{eqnarray}

\subsection{The RS model}

In the RS model, the corresponding metric is
\begin{eqnarray}
ds^2&=&e^{2\phi(\xi)}\eta_{\mu\nu}dx^\mu dx^\nu+
d\xi^2\nonumber \\
\phi(\xi)&=&-\kappa\xi
\end{eqnarray}
where $0\leq\xi\leq\pi r_c$ is the coordinate for an extra
dimension and $r_c$ is the "compactification radius" of the extra
dimension \cite{RS}. The Klein-Gorden equation reads
\begin{eqnarray}
{1\over
\sqrt{-g}}\partial_A(\sqrt{-g}g^{AB}\partial_B\Psi)-m^2\Psi&=&0.
\end{eqnarray}
Substituting the metric into the field equation, one has
\begin{eqnarray}
-4\kappa{\partial\Psi\over \partial
\xi}-e^{2\kappa\xi}({\partial\over
\partial t})^2\Psi+e^{2\kappa\xi}
({\partial\over
\partial \vec{x}})^2\Psi+({\partial\over
\partial \xi})^2\Psi-m^2\Psi&=&0.\label{formula1}
\end{eqnarray}
Similar to the flat space-time case, decomposing the wavefunction
$\psi$ into a product form
\begin{eqnarray}
\Psi&=&e^{i\vec{k}\cdot\vec{x}}e^{-2\phi(\xi)}\psi(\xi,t),
\end{eqnarray}
and substituting it into eq.(\ref{formula1}), we eventually obtain
the equation which only contains differentiation of $\psi$ with
respect to the fifth dimension $\xi$ and time $t$,
\begin{eqnarray}
\left[({\partial\over
\partial\xi})^2-e^{2\kappa\xi}({\partial\over \partial t})^2-(4\kappa^2+m^2+\vec{k}^2e^{2\kappa\xi})\right]\psi(\xi,t)
&=&0\label{KG}.
\end{eqnarray}
Then with the non-relativistic approximation,
\begin{eqnarray}
\psi(\xi,t)&=&\varphi(\xi,t)e^{-imt}\nonumber \\
-({\partial \over \partial t})^2\psi&\simeq&\left[2im{\partial
\varphi\over \partial t}+m^2\varphi\right]e^{-imt}
\end{eqnarray}
we get
\begin{eqnarray}
2me^{2\kappa\xi}i{\partial \varphi\over \partial t}+({\partial
\over
\partial
\xi})^2\varphi+\left(e^{2\kappa\xi}m^2-4\kappa^2-m^2-\vec{k}^2e^{2\kappa\xi}\right)\varphi&=&0.
\end{eqnarray}
Further, we can rewrite the above equation as
\begin{eqnarray}
({\partial \over \partial
\xi})^2\varphi&=&\left[(\vec{k}^2-m^2-2mE)e^{2\kappa\xi}+m^2+4\kappa^2\right]\varphi.\label{RS1}
\end{eqnarray}

Considering two $\delta$ barriers at the two brane surfaces, we
finally arrive at what we want to have
\begin{eqnarray}
-{1\over 2m}({\partial \over \partial \xi})^2\varphi+\left[{1\over
2m}(\vec{k}^2-m^2)e^{2\kappa\xi}+{1\over
2m}(m^2+4\kappa^2)+a\delta(\xi)+a\delta(\xi-\pi
r_c)\right]\varphi&=&Ee^{2\kappa\xi}\varphi.\label{RS2}
\end{eqnarray}

At the anti-world brane, the solution of eq.(\ref{RS2}) is:
\begin{eqnarray}
\varphi(\xi)&=&e^{i\alpha(\xi)}+R_1e^{-i\alpha
(\xi)}\;\;\;\;\;\;\;\;\;\;\xi<0,\nonumber \\
\varphi(\xi)&=&S_1e^{i\alpha(\xi)}+R_2e^{-i\alpha(\xi)}\;\;\;\;\;\;\xi>0,
\end{eqnarray}
and at our brane (matter), the solution of eq.(\ref{RS2}) is:
\begin{eqnarray}
\varphi(\xi)&=&S_1e^{i\alpha(\xi)}+R_2e^{-i\alpha(\xi)}\;\;\;\;\;\;\xi<\pi r_c,\nonumber \\
\varphi(\xi)&=&S_2e^{i\alpha(\xi)}\;\;\;\;\;\;\;\;\;\;\;\;\;\;\;\;\;\;\;\;\;\;\;\;\;\xi>\pi
r_c.
\end{eqnarray}
where $\alpha(\xi)$ is the eigenvalue of eq.(\ref{RS1}):
$\alpha(\xi)={\sqrt{2mE+m^2-\vec{k}^2}\over \kappa}e^{\kappa\xi}$.

With the same boundary conditions which were depicted for the flat
space-time case, we can find the barrier penetration rate
$T=|S_2|^2$ as
\begin{eqnarray}
T&=&|S_2|^2=1/\left\{\left[({ma\over \kappa})^2{1\over
\beta_1\beta_2}(\cos(2(\beta_2-\beta_1))-1)+1\right]^2+\left[({ma\over
\kappa})^2{1\over \beta_1\beta_2}\sin(2(\beta_2-\beta_1))+{ma\over
\kappa}({1\over \beta_1}+{1\over
\beta_2})\right]^2\right\},\nonumber
\\
\beta_1&=&{\sqrt{2mE+m^2-\vec{k}^2}\over \kappa},\nonumber \\
\beta_2&=&{\sqrt{2mE+m^2-\vec{k}^2}\over \kappa}e^{\pi \kappa
r_c}.
\end{eqnarray}
This expression of transition rate is different from that for the
flat space-time case, some details would be manifested in the
numerical results and the following figures.

\subsection{The evolution of the two branes in RS model}
The key point concerning the RS model is whether the evolution of
the two-brane structure coincides with the present astronomical
observation. To investigate the evolution, one needs to solve the
five-dimensional Einstein's equations for the "compactification
radius" $r_c$ at any time\cite{Einstein}. We rewrite the eq.(20)
into a different form by assuming $\xi=r_c(t)\widetilde{\xi}$:
\begin{eqnarray}
ds^2&=&(-e^{-2\kappa
r_c(t)\widetilde{\xi}}+\widetilde{\xi}^2\dot{r_c}^2)dt^2+e^{-2\kappa
r_c(t)\widetilde{\xi}}d\vec{x}^2+r_c(t)^2d\widetilde{\xi}^2+
2\widetilde{\xi}r_c(t)\dot{r_c}d\widetilde{\xi}dt
\end{eqnarray}

The five-dimensional Einstein's equations are
\begin{eqnarray}
G_{AB}\equiv R_{AB}-{1\over 2}Rg_{AB}=\widetilde{\kappa}^2T_{AB}
\end{eqnarray}
where $R_{AB}$ is the five-dimensional Ricci tensor,
$R=g^{AB}R_{AB}$ the scalar curvature and the constant
$\widetilde{\kappa}$ is related to the five-dimensional Newton's
constant with $\widetilde{\kappa}^2=8\pi G_{(5)}$\cite{Einstein1}.
The right hand term $T_{AB}$ is the energy-momentum tensor.

Inserting the metric in eq.(32) into the Einstein equations, we
can obtain the non-vanishing components of the Einstein tensor
$G_{AB}$ which includes a derivative of $r_c$ with respect to time
t. In principle, it is a self-consistent differential equation
group and would be extremely difficult to solve, but with a
reasonable approximation, one can be priori set the
energy-momentum tensor for the bulk matter and the matter content
on the branes, the differential equations can be solved. Then we
would be able to obtain $r_c$ at any time. In practice, because
the five-dimensional differential equations for $r_c(t)$ and the
junction conditions are too complex, that even with the
assumption, we are unable at present to get a solution, no matter
analytical or numerical. In order to discuss the physics picture,
let us take an extreme simplification. In eq. (A1) which is
presented in the appendix, we assume that $\dot r_c$ is small, so
that we can neglect the terms with higher powers of $\dot r_c$ and
only keep $\dot r_c^2$ terms on the left-side of eq.(A1). The
right-side of the energy tensor $T_{00}$ reflects a competition
between the Casimir repulsion and gravitational attraction. As
shown in previous subsection, at very early universe the Casimir
repulsion dominates over the gravitational attraction between the
matter on the two branes, $T_{00}$ is positive. If we can
approximately set $T_{00}$ as a constant, the equation is simple
and the solution is
\textcolor[rgb]{1.00,0.00,0.00}{$r_c(t)\sim\exp(\alpha t)$} where
$\alpha$ is a constant related to the $T_{00}$ and other
parameters. It is an exponentially increasing function, namely the
two branes would separate by the repulsive force. However, as
shown above, $T_{00}$ is not a constant, and when $r_c$ reaches a
certain value, the gravitational attraction becomes stronger, then
the separation would slow down till completely stops when the two
forces balance each other. Indeed, a complete solution for the
evolution process is beyond the scope of the work and our present
ability, we will pursue this topic in our future studies.

\section{Numerical Result for phenomenology}

Here for the numerical computations of the flux of antimatter to
be detected at the AMS, we include all the necessary input
parameters which are directly adopted from the concerned published
literatures\cite{RS,data0}.
\begin{eqnarray}
T&=&\bar\rho_{anti-helium}/\rho_{helium}
\end{eqnarray}
where $\bar\rho_{anti-helium}$ is the mass density of the
anti-helium particles which overcome the brane barriers to transit
into our brane, $m=m_{He}=4$ GeV, the dispersive velocity of
helium is $v_{He}=1000$ km/s and $\kappa r_c=12$. The
"compactification radius" of the extra dimension $r_c$ and factor
$a$ in the $\delta$ potential are regarded as free parameters.
\begin{figure}[!htb]
\begin{center}
\begin{tabular}{cc}
\includegraphics[width=9cm]{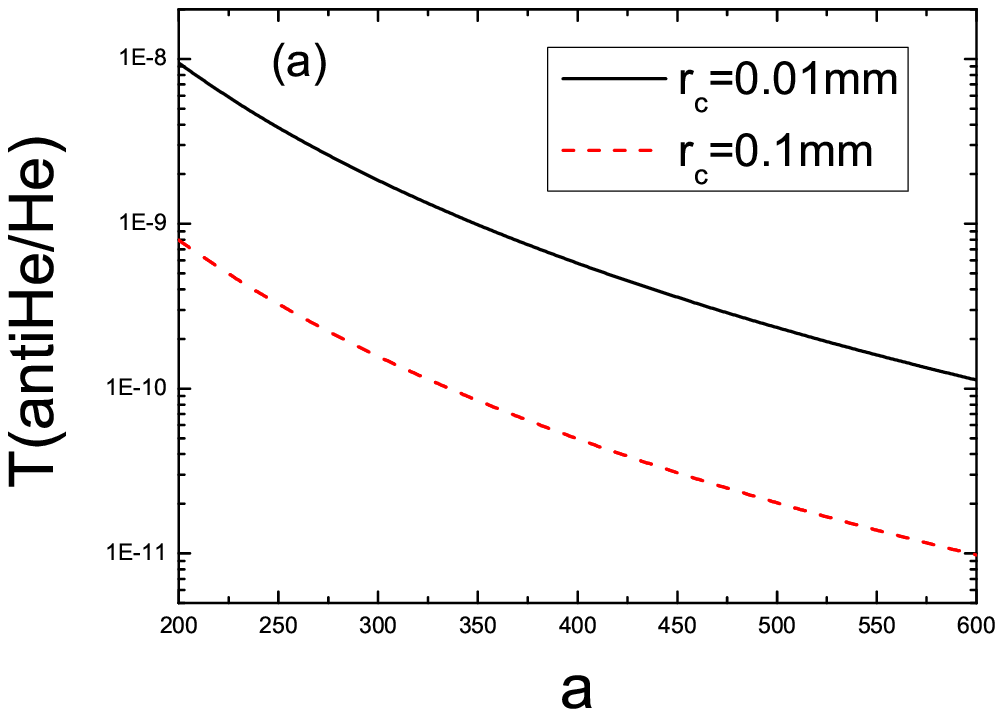}
\includegraphics[width=9cm]{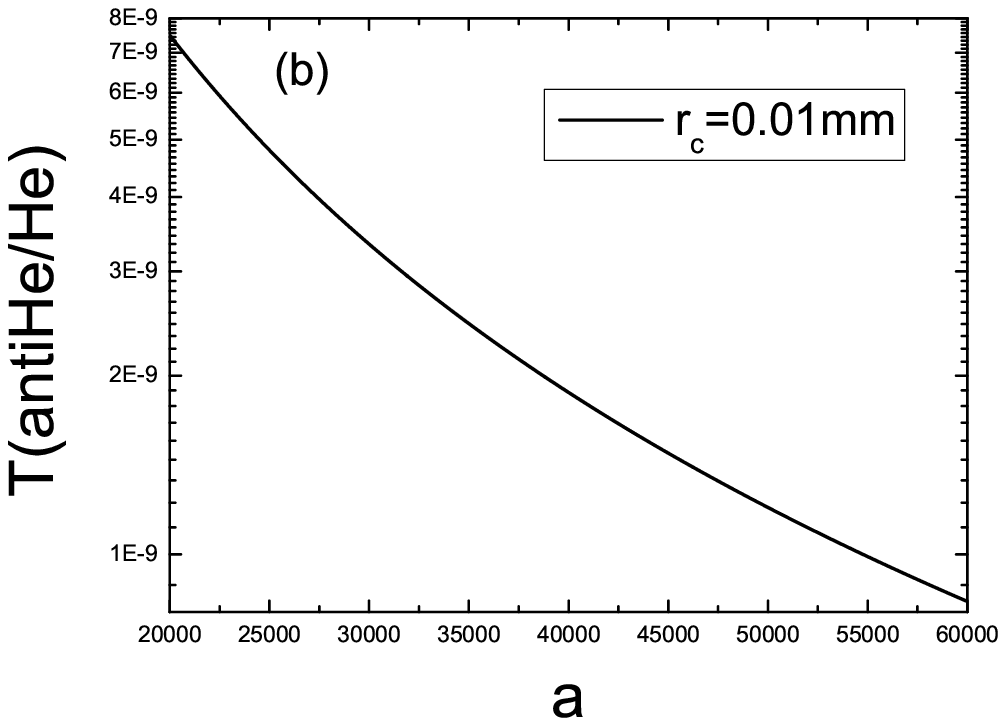}
\end{tabular}
\end{center}
\label{figure} \caption{The anti-He/He flux-ratio (a) for the flat
space-time model and (b) for the RS model.} \label{fig1}
\end{figure}
\vspace{0.6cm}

In Fig. \ref{fig1}, we show the ratio of the anti-helium flux over
helium flux in our three-dimension space, which can be detected by
our detector on the earth or AMS. In this work, we are working on
the naive flat spacetime and the RS-I model, the physics condition
we set should determine a bound on $r_c$, while similar bounds may
be gained by solving the hierarchy problem\cite{RS} or the minimum
condition of the effective potential\cite{Wise}. Here we choose
$r_c=0.01$ and 0.1 mm, which is consistent with present data on
gravity. It is noted that the ratio drops very fast as the
potential strength $a$ increases for the naive flat space-time
model, but not so abruptly for the RS model. The flux ratio
decreases very fast as the distance $r_c$ increases for the naive
flat space-time model, but almost does not vary for the RS model.
Let us roughly estimate the order of magnitude of the surface
potential. It is of order $a/r_c\sim 2^{-8} a$ MeV, and as $a\sim
1000$, it is a few hundreds of eV. It seems reasonable.

\begin{figure}[!htb]
\begin{center}
\begin{tabular}{cc}
\includegraphics[width=9cm]{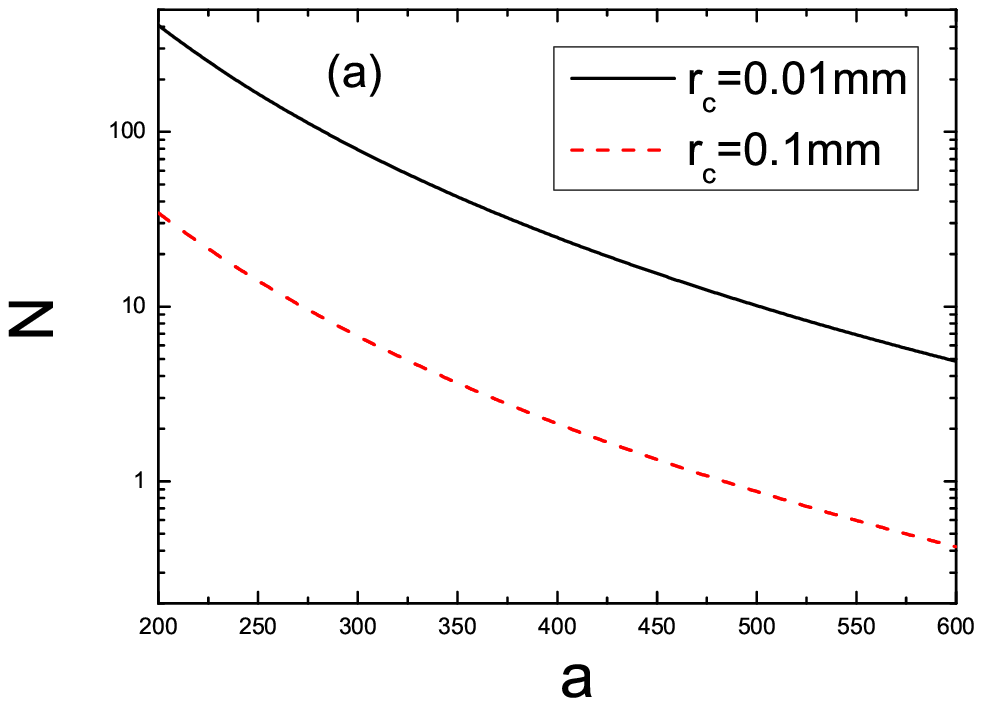}
\includegraphics[width=9cm]{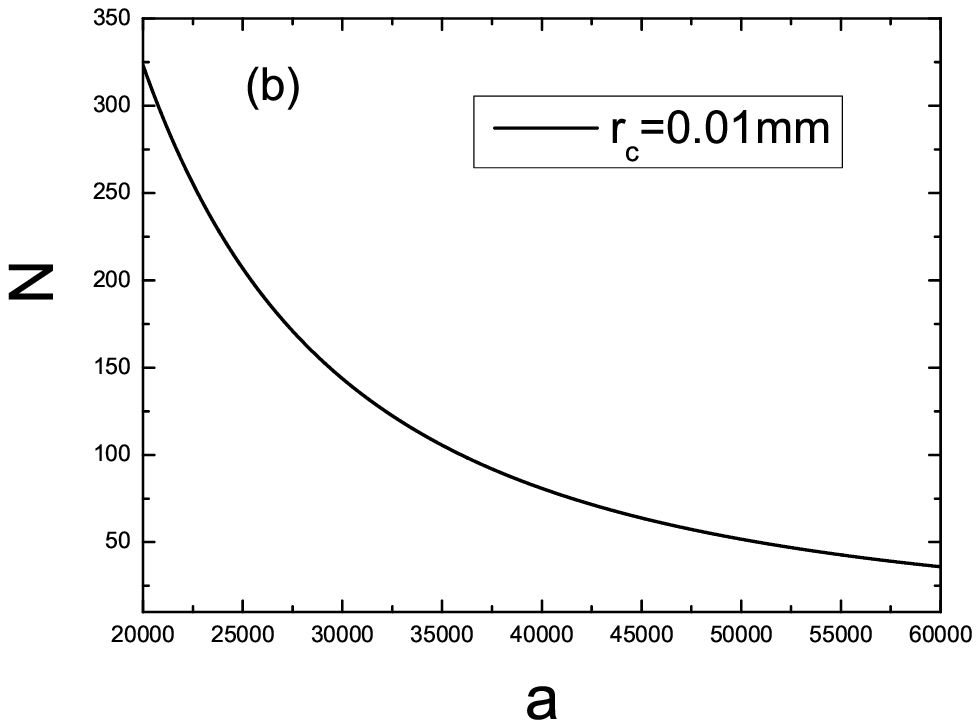}
\end{tabular}
\end{center}
\label{figure} \caption{The number of anti-helium particles which
can be detected by the AMS in one year (a)  for the flat
space-time model (b) for the RS model} \label{fig2}
\end{figure}

The numbers of anti-helium particles which can be detected by the
AMS should be
\begin{equation} N={\Delta \Omega\over 4\pi}{\bar\rho_{anti-helium}\over m} |{\bf v}| \Delta S\Delta t\times
f,
\end{equation}
where the factor $\Delta \Omega/ 4\pi$ is from the random
direction of the flux, $\bar\rho_{anti-helium}=\rho_{helium}\times
T$ and $\rho_{helium}=23\%\rho_b$, $\rho_b=0.042\rho_c$,
$\rho_c=1.87837\times10^{-29}h^2$gcm$^{-3}$, $h=0.73$\cite{data0},
m is the mass of single helium particle, $|{\bf v}|=v_{He}$ is the
average velocity of the anti-helium, $\Delta t$ is the duration of
the detection which we take as one year, $\Delta S$ is the area of
AMS with $\Delta \Omega \Delta S=0.65$ sr $m^2$\cite{AMS2} and $f$
is the detection efficiency of detecting such anti-helium
particles by the AMS which we take as 100$\%$. In the
\cite{dispersive}, the authors decide that the average dispersion
velocity of dark matter particles is within a range of 600 to 1000
km/s, and we just adopt the maximal value as a reasonable
approximation for $v_{He}$. Obviously the theoretical prediction
of $N$ depends on the model and the concerned parameters such as
$r_c$ and $a$. Later, with a few typical parameter sets, we
tabulate the number of anti-helium particles which may be detected
by AMS in Fig. \ref{fig2}.

\begin{table}[h]
\caption{The number of anti-helium particles which can be detected
by the AMS in one year for the flat space-time metric.}
\begin{center}
\begin{tabular}{|c|c|c|c|c|c|c|c|c|c|}
  \hline
  N & $a=200$ & $a=250$ & $a=300$ & $a=350$ & $a=400$ & $a=450$ & $a=500$ & $a=550$ & $a=600$\\
  \hline
  $r_c=0.01$ mm & $408$ & $165$ & $76$ & $43$ & $25$ & $15$ & $10$ & $7$ & $5$\\
  \hline
  $r_c=0.1$ mm & $34$ & $14$ & $7$ & $4$ & $2$ & $1$ & $0$ & $0$ & $0$\\
  \hline
\end{tabular}
\end{center}\label{flambdab}
\end{table}

\begin{table}[h]
\caption{The number of anti-helium particles which can be detected
by the AMS in one year for the R-S metric.}
\begin{center}
\begin{tabular}{|c|c|c|c|c|c|c|c|c|c|}
  \hline
  N & $a=20000$ & $a=25000$ & $a=30000$ & $a=35000$ & $a=40000$ & $a=45000$ & $a=50000$ & $a=55000$ & $a=60000$\\
  \hline
  $r_c=0.01$ mm & $323$ & $207$ & $144$ & $106$ & $81$ & $64$ & $52$ & $43$ & $36$\\
  \hline
  $r_c=0.1$ mm & $323$ & $207$ & $144$ & $106$ & $81$ & $64$ & $52$ & $43$ & $36$\\
  \hline
\end{tabular}
\end{center}\label{flambdab}
\end{table}

\section{Discussion and Conclusion}

In this work we propose a possible physical picture to interpret
the matter-antimatter asymmetry observed in our universe. We
suppose that at the early epoch of the universe evolution, maybe
after the inflation stage, the CP symmetry is spontaneously
broken, and the two branes correspond to the two ground states of
CP. Thus the matter and antimatter were separated onto two
different branes. A complex scalar field which only applies to the
extra dimension, is introduced to be responsible for the
spontaneous CP symmetry breaking and the CP phase-dependent vacuum
expectation values can be different at the two branes. The scalar
field existing between the two branes, causes a Casimir force
which repels the two branes away from each other.  The two branes
would attract each other via gravitational force. Preventing the
two branes to collide and matter-antimatter annihilate, the scalar
field which obeys anti-periodic boundary conditions on the two
branes provides a repulsive force to opposes the gravitational
attraction. For smaller distance between two branes, as shown in
the text, the Casimir repulsive force is stronger than the
gravitational attraction and the consequence is that the two
branes would be pushed away from each other till sometime which
would be much later than today, the two forces are balanced and an
equilibrium is reached (roughly, the separation would be a few
hundreds of km). With extra dimensions, the $1/r^2$ Newton's
gravitational law must be modified \cite{G45_1,G45_2} as shown in
the form of eq. (\ref{newton}). Since today one does not observe
any deviation from the $1/r^2$ law at the macroscopic scale, he
must consider that $r_c$ is sufficiently small, such as less than
0.1 mm. In this work, we take $r_c=0.1$ and 0.01 mm, of course it
is only an illustration. In many, many years, when the two branes
are separated very far by meters, the observational gravitation
law will deviate from $1/r^2$ form. If one needs to find the
evolution of the brane world, namely how the two branes are
separated from initial $r_c\approx 0$ to the present value, he
must solve the 5-dimensional time-dependent Einstein equation, but
it is beyond the scope of the present work and we will not discuss
the evolution process here.

As discussed by many authors, generally the matter (antimatter)
and gauge bosons are forbidden to enter the fifth dimension except
the gravitational force lines, to realize the picture, we suggest
that there is a barrier on the edge of the brane in analog to the
surface tension of water membrane. We use a simple delta function
to describe the barrier. Like the picture for Hawking radiation of
black holes, the quantum effects may cause a quantum tunnelling of
the matter and antimatter from one brane to another.

The antimatter which reside on another brane would have a
probability to transit into our brane with matter only. The flux
depends on the barrier strength and may be detected by the
detector on earth. The AMS would be an ideal apparatus to do the
job. According to the preliminary results of AMS on the antimatter
flux\cite{AMS}, we can estimate the brane-barrier strength. In
this work, we consider two popular models, the naive flat
space-time model and the R-S models, to carry out the
calculations. We find that their results about antimatter flux are
quite different as shown in Fig. \ref{fig1}.

This picture is indeed somehow ad hoc and speculative, but
provides a possible interpretation for the matter-antimatter
asymmetry observed in our universe, and suggests an existence of
the antimatter flux which can be detected by AMS. There are indeed
a few adjustable parameters in the picture which cannot be
determined from the first principle so far and need to be fixed by
the measurements of AMS. We are eagerly waiting for the
measurement results of AMS because they may tell us much more
information about the universe
and also probe our proposal.\\

\noindent Acknowledgement:

We benefit greatly from very stimulating and active discussions
with Xiao-Gang He, indeed some initiative ideas are produced
during such conversations. We also thank H.S. Chen for helpful
discussions and introduction about new progress on the AMS
project. Discussion with Liu Zhao is also helpful and fruitful.
This work is partly supported by the National Natural Science
Foundation of China (NNSFC), No.10475042.

\vspace{1cm}

\noindent{\bf Appendix: The five-dimensional Einstein equations including derivative with respect to time}\\

\begin{eqnarray}
&&3\kappa(9\widetilde{\xi}^5\dot{r_c}^6-3\kappa\dot{r_c}^6r_c\widetilde{\xi}^6+
2\kappa r_c e^{-6\kappa r_c\widetilde{\xi}}+e^{-4\kappa
r_c\widetilde{\xi}}\widetilde{\xi}\dot{r_c}^2+4e^{-2\kappa
r_c\widetilde{\xi}}\widetilde{\xi}^3r_c
\ddot{r_c}\dot{r_c}^2+9\widetilde{\xi}^2\dot{r_c}^2r_c\kappa
e^{-4\kappa
r_c\widetilde{\xi}}+4\widetilde{\xi}^4\dot{r_c}^4r_c\kappa
e^{-2\kappa r_c\widetilde{\xi}}\nonumber
\\
&&-2\widetilde{\xi}^3\dot{r_c}^4e^{-2\kappa
r_c\widetilde{\xi}})/(r_c(e^{-2\kappa
r_c\widetilde{\xi}}+3\widetilde{\xi}^2\dot{r_c}^2)^2)\nonumber
=\widetilde{\kappa}^2T_{00}, \hspace{3cm} (\text{A}1)\nonumber
\\
&&-e^{-2\kappa r_c\widetilde{\xi}}(-2e^{-2\kappa
r_c\widetilde{\xi}}\kappa
r_c^2\widetilde{\xi}\ddot{r_c}+29e^{-2\kappa
r_c\widetilde{\xi}}\kappa^2r_c^2\widetilde{\xi}^2\dot{r_c}^2+
18\widetilde{\xi}^4\dot{r_c}^4\kappa^2r_c^2-12\widetilde{\xi}^3\dot{r_c}^4\kappa
r_c+6\kappa^2r_c^2e^{-4\kappa r_c\widetilde{\xi}}+e^{-2\kappa
r_c\widetilde{\xi}}r_c\ddot{r_c}\nonumber \\
&&+e^{-2\kappa r_c\widetilde{\xi}}\dot{r_c}^2+5\kappa
\dot{r_c}^2\kappa e^{-2\kappa
r_c\widetilde{\xi}}r_c)/(r_c^2(e^{-2\kappa
r_c\widetilde{\xi}}+3\widetilde{\xi}^2\dot{r_c}^2)^2)\nonumber
\\
&&=\widetilde{\kappa}^2T_{11}=\widetilde{\kappa}^2T_{22}=\widetilde{\kappa}^2T_{33},
\hspace{7cm}(\text{A}2)\nonumber
\\
&&-3\kappa r_c(2e^{-4\kappa r_c\widetilde{\xi}}\kappa
r_c+8e^{-2\kappa r_c\widetilde{\xi}}\kappa
r_c\widetilde{\xi}^2\dot{r_c}^2-
\widetilde{\xi}\dot{r_c}^2e^{-2\kappa
r_c\widetilde{\xi}}+3\widetilde{\xi}^4\dot{r_c}^4\kappa
r_c-9\widetilde{\xi}^3\dot{r_c}^4-e^{-2\kappa
r_c\widetilde{\xi}}\widetilde{\xi}r_c\ddot{r_c})/(e^{-2\kappa
r_c\widetilde{\xi}}+3\widetilde{\xi}^2\dot{r_c}^2)^2\nonumber
\\
&&=\widetilde{\kappa}^2T_{44},\hspace{10cm}(\text{A}3) \nonumber
\\
&&-3\dot{r_c}\widetilde{\xi}\kappa(e^{-2\kappa
r_c\widetilde{\xi}}\widetilde{\xi}\dot{r_c}^2-9\widetilde{\xi}^3\dot{r_c}^4+
4\kappa r_ce^{-4\kappa
r_c\widetilde{\xi}}+19\widetilde{\xi}^2\dot{r_c}^2r_c\kappa
e^{-2\kappa r_c\widetilde{\xi}}+15\kappa \dot{r_c}^4r_c
\widetilde{\xi}^4-2e^{-2\kappa
r_c\widetilde{\xi}}\widetilde{\xi}r_c\ddot{r_c})/(e^{-2\kappa
r_c\widetilde{\xi}}+3\widetilde{\xi}^2\dot{r_c}^2)^2\nonumber \\
&&=0,\hspace{11cm}(\text{A}4)\nonumber
\end{eqnarray}
where $T_{00}, T_{11}, T_{22}, T_{33}, T_{44}$ are the components
of energy-momentum tensor $T_{AB}$ of the bulk matter and the
matter content in the brane, which expressions can be found in
Ref.\cite{Einstein1}.

\end{document}